\documentclass[journal]{IEEEtran}
\usepackage{graphicx}
\usepackage[pdftex]{color}

\usepackage[]{amssymb}
\usepackage[]{textgreek}

\usepackage[alsoload=synchem]{siunitx}
\DeclareSIUnit\pixel{pixel}
\DeclareSIUnit\electron{e^{-}}
\DeclareSIUnit\adu{ADU}

\ifCLASSINFOpdf
\else
\fi
\usepackage[cmex10]{mathtools}

\usepackage{physics}

\begin{document}
\title{Robust Pixel Gain Calibration with Limited Statistics}%

\author{G.~Blaj*,~\IEEEmembership{Member,~IEEE},
        G.~Haller,~\IEEEmembership{Member,~IEEE},
        C.~Kenney,~\IEEEmembership{Member,~IEEE}% <-this % stops a space
%\thanks{Manuscript received.... SLAC-PUB-xxxxx.}
\thanks{G.~Blaj, G.~Haller, and C.~Kenney are with SLAC National Accelerator Laboratory, Menlo Park, CA 94025.}% <-this % stops a space
\thanks{* Corresponding author: blaj@slac.stanford.edu}%
}


\markboth{}%
{Shell \MakeLowercase{\textit{et al.}}: Bare Demo of IEEEtran.cls for Journals}



\maketitle

%\begin{abstract}
%\boldmath

%\end{abstract}

%\begin{IEEEkeywords}
%Hybrid pixel detectors, gain calibration, laboratory sources
%\end{IEEEkeywords}

% \fi
\IEEEpeerreviewmaketitle

\section{Introduction}
Pixel detectors typically display pixel-to-pixel gain variation of a few percent which result in reduced spectroscopic performance \cite{blaj2019deadtime}. Integrating pixel detectors provide spectroscopic information directly, while photon counting detectors through threshold scans. Gain maps are used to rectify pixel-to-pixel gain variation.

Ideally, the pixel gain is calibrated with monochromatic radiation (e.g., from a synchrotron or free electron laser beamline, or laboratory x-ray sources with monochromators). However, beam time is usually valuable, and monochromatic sources are not always available \cite{klavckova2019characterization}, while x-ray tubes and radioactive sources produce complex spectra.

For calibration of ePix100a cameras, we used a Mo x-ray tube with Zr and Al filters (to optimize the Mo K\textalpha{ }line) and acquired limited statistics (\numrange{50}{250} photons), sampled from a complex spectrum (where accurate fitting of a peak with charge sharing would require $\approx$\num{1000} photons per pixel \cite{blaj2016xray}).

We have developed a calibration method which relies on cross-correlating histograms of many pixel pairs and obtaining large sets of relative shifts. These were subsequently used to calculate absolute pixel shifts and corresponding pixel gains.

We demonstrate that this method yields stable gain calibration maps with an order of magnitude less statistics than required by typical approaches. Finally, we demonstrate the accuracy of the method by comparing with gain maps obtained with good statistics and monochromatic radiation at a synchrotron beamline.

The robust gain calibration method presented here can be used for any pixel detector with minimal effort and assumptions on calibration spectrum. It is particularly useful when the quantity and/or quality of calibration data is limited and repeated measurements could be difficult (e.g., previous experiments at synchrotron or free electron laser sources, FEL).

\section{Methods}

\begin{figure}[!t]
\centering
\includegraphics[width=\columnwidth]{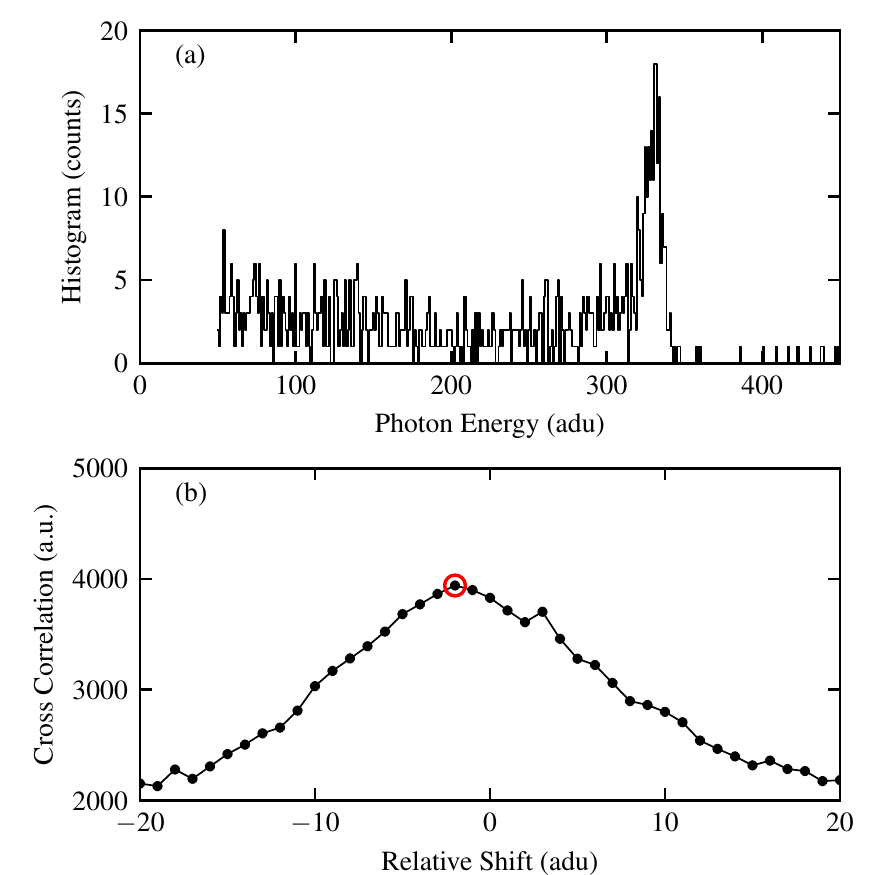}
% where an .eps filename suffix will be assumed under latex, 
% and a .pdf suffix will be assumed for pdflatex; or what has been declared
% via \DeclareGraphicsExtensions.
\caption{Determining peak positions in statistics limited spectra is nontrivial; (a)~shows a spectrum collected by one pixel; the spectrum displays significant charge sharing, a low energy shoulder, and limited statistics (on average, \num{240}~photons per pixel near the Mo~K\textalpha{ }peak); (b)~calculating the cross correlation of two pixel spectra and finding the maximum value yields the most likely shift between the two pixel spectra (indicated by the red circle).}
\label{fig1}
\end{figure}

\begin{figure}[!t]
\centering
\includegraphics[width=\columnwidth]{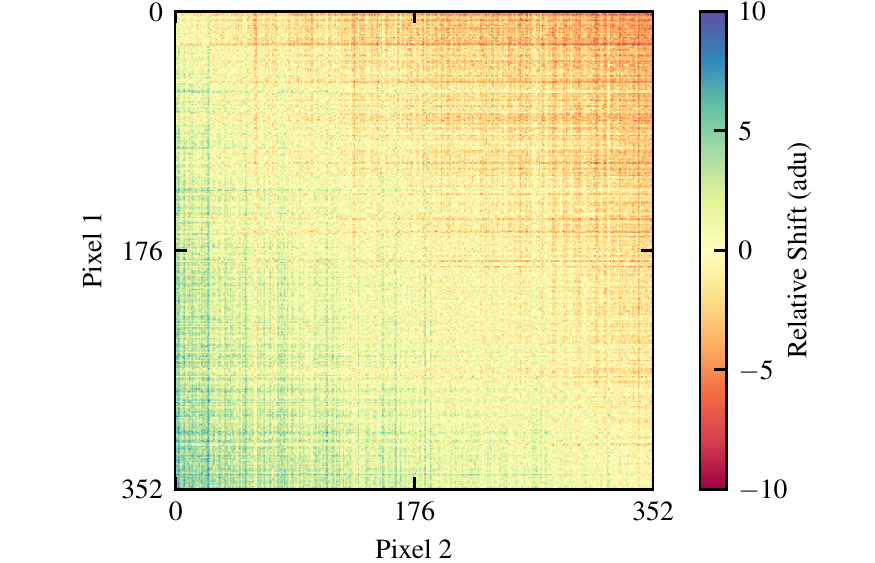}
% where an .eps filename suffix will be assumed under latex, 
% and a .pdf suffix will be assumed for pdflatex; or what has been declared
% via \DeclareGraphicsExtensions.
\caption{The shift matrix contains relative delays for all possible pairs of pixels; individual delays could be noisy, however, averaging large numbers yields good shift estimates; we show the shift matrix for pixels in one column.}
\label{fig2}
\end{figure}

We used an x-ray tube with Mo anode and Zr and Al filters to emphasize the Mo~K\textalpha{ }peak \cite{blaj2016xray}. We acquired two data sets of $\approx$\num{33000}~frames each, with the x-ray tube operated at \SI{45}{\kilo\volt} and \SI{50}{\kilo\volt}, respectively. We collected histograms of ``single pixel events'' for each pixel (i.e., all 8 neighboring pixels within noise, $\le 4 \sigma$), yielding an average of \num{240} photons per pixel. Fig.~\ref{fig1}(a) shows an example of pixel spectrum with limited statistics. The peak position obtained through fitting or centroiding is relatively noisy \cite{blaj2016xray}.

A change in pixel gain results in an apparent shift of (calibration) peak position. While determining the peak shift is nontrivial with limited statistics, we developed a robust approach for spectra with a sharp transition (i.e., from a quasi-monochromatic source and/or an absorption edge filter). For each pair of pixel spectra we estimate the relative peak shift by calculating the cross correlation, finding its maximum value and corresponding delay as in Fig.~\ref{fig1}(b). This delay is the relative shift of the two pixel spectra.

The relative shift pairs are saved in a ``shift matrix'' of size $N\times N$ (with $N$ the number of pixels), see Fig.~\ref{fig2}. With limited statistics, relative shifts of individual pairs could be imprecise, however, the column average of the shift matrix yields a robust and accurate estimate of the absolute pixel shift\footnote{Demonstration omitted for brevity.}.

The algorithm complexity is $O(N^2)$; for large numbers of pixels we can optimize the run time by dividing the problem and calculating (1)~shifts for pixels within each column, and (2)~shifts of individual columns; their sum yields the absolute pixel $i$ shift $\delta_i$ and the pixel gain as $g_i=1+\delta_i/E_0$ where $E_0$ is the peak position.

\section{Results}

\begin{figure}[!t]
\centering
\includegraphics[width=\columnwidth]{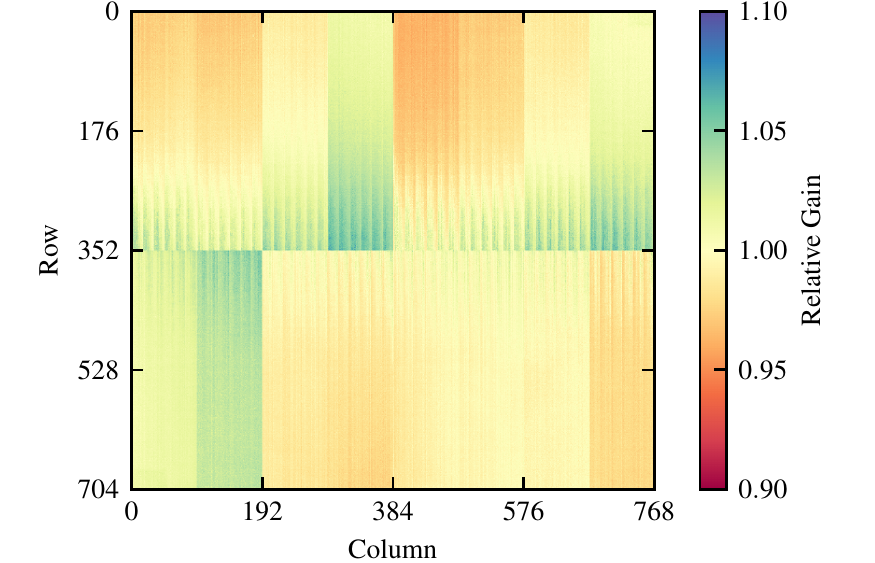}
% where an .eps filename suffix will be assumed under latex, 
% and a .pdf suffix will be assumed for pdflatex; or what has been declared
% via \DeclareGraphicsExtensions.
\caption{Converting pixel shifts to gains yields a robust detector gain map, despite limited statistics and spectrum quality; in this ePix100a camera, \SI{99.2}{\percent} of the pixels have a relative gain in the \numrange{0.95}{1.05} range.}
\label{fig4}
\end{figure}

\begin{figure}[!t]
\centering
\includegraphics[width=\columnwidth]{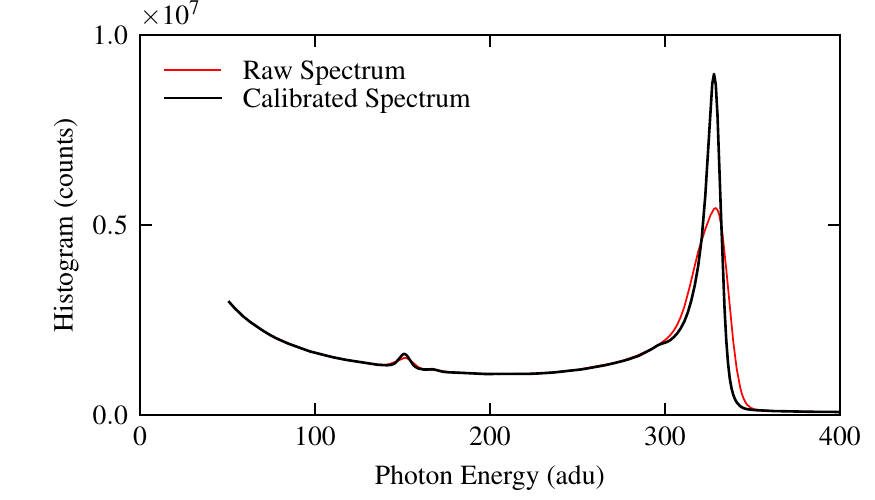}
% where an .eps filename suffix will be assumed under latex, 
% and a .pdf suffix will be assumed for pdflatex; or what has been declared
% via \DeclareGraphicsExtensions.
\caption{Spectrum integrated over all pixels; the thin red line shows the raw spectrum and the thick black line shows the spectrum after gain calibration; the spectroscopic performance improves significantly by using a gain map.}
\label{fig5}
\end{figure}

Fig.~\ref{fig4} shows the robust gain map obtained for an ePix100a camera, showing good uniformity (\SI{99.2}{\percent} of the pixel gains in the \numrange{0.95}{1.05} range). Fig.~\ref{fig5} shows the aggregated spectrum of all pixels, either raw (thin red line) or after gain calibration (thick black line). The spectroscopic performance is significantly improved by using a gain map \cite{blaj2018performance,blaj2018hammerhead}.

\begin{figure}[!t]
\centering
\includegraphics[width=\columnwidth]{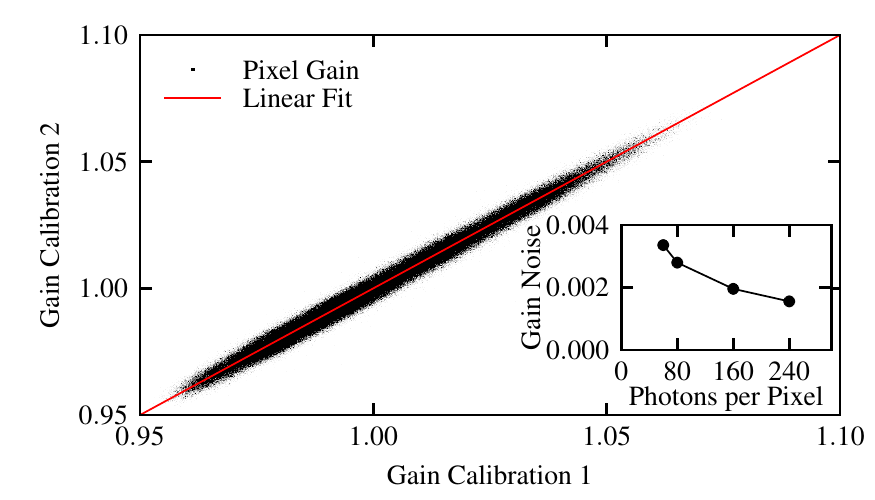}
% where an .eps filename suffix will be assumed under latex, 
% and a .pdf suffix will be assumed for pdflatex; or what has been declared
% via \DeclareGraphicsExtensions.
\caption{Comparing gain maps obtained from two independent, statistics limited measurements; each black dot corresponds to one pixel with the two coordinates representing the two independent gains; the red line is a linear fit with a slope of $\num{1.000}~\pm~\num{3E-6}$, a $R^2=0.987$ and residual gain noise \SI{0.16}{\percent} for the complete data set (\num{240} photons). The inset shows the gain noise dependence on statistics; even for a reduced set with \num{60} photons per pixel, the gain noise is low (\SI{0.34}{\percent}).}
\label{fig6}
\end{figure}

To test the robustness of this method, we applied it to two independent data sets obtained in different conditions. The results are shown in Fig.~\ref{fig6}, with each black point indicating two independent gain estimations of one pixel. A linear regression yields a slope $a=\num{1.000}~\pm~\num{3e-6}$, with $R^2=\num{0.987}$ and residual gain noise \SI{0.16}{\percent}. This corresponds to an error due to gain calibration of \num{0.51}~adu at the Mo~K\textalpha{ }peak, demonstrating good repeatability of the robust gain calibration with different statistics limited data sets. Finally, reducing the calibration data from \num{240} to \num{60} photons per pixel results in a relatively small increase of gain noise to \SI{0.34}{\percent}.

\section{Conclusion}
We presented a robust pixel gain calibration, demonstrating good performance and reproducibility with limited spectrum quality and statistics (an order of magnitude less than required by other methods). This method is relatively fast, with a Python program calibrating \num{0.5}~megapixel ePix100a cameras in about \SI{16}{\minute} on a laptop with \SI{2.9}{\giga\hertz} CPU. It is also simple to implement, requiring minimal assumptions on spectrum shape.

The robust gain calibration method could be particularly useful when the quantity and/or quality of calibration data is limited and repeated measurements could be difficult (e.g., previous experimental data at synchrotron or FEL sources).

% or


%\appendices
%Some text for the appendix.

% Acknowledgement
\section*{Acknowledgement}
Use of the Linac Coherent Light Source (LCLS) and Stanford Synchrotron Radiation Lightsource (SSRL), SLAC National Accelerator Laboratory, is supported by the U.S. Department of Energy, Office of Science, Office of Basic Energy Sciences under Contract No. DE-AC02-76SF00515. Publication number SLAC-PUB-17459.


\begin{thebibliography}{1}
\providecommand{\url}[1]{#1}
\csname url@samestyle\endcsname
\providecommand{\newblock}{\relax}
\providecommand{\bibinfo}[2]{#2}
\providecommand{\BIBentrySTDinterwordspacing}{\spaceskip=0pt\relax}
\providecommand{\BIBentryALTinterwordstretchfactor}{4}
\providecommand{\BIBentryALTinterwordspacing}{\spaceskip=\fontdimen2\font plus
\BIBentryALTinterwordstretchfactor\fontdimen3\font minus
  \fontdimen4\font\relax}
\providecommand{\BIBforeignlanguage}[2]{{%
\expandafter\ifx\csname l@#1\endcsname\relax
\typeout{** WARNING: IEEEtran.bst: No hyphenation pattern has been}%
\typeout{** loaded for the language `#1'. Using the pattern for}%
\typeout{** the default language instead.}%
\else
\language=\csname l@#1\endcsname
\fi
#2}}
\providecommand{\BIBdecl}{\relax}
\BIBdecl

\bibitem{blaj2019deadtime}
\BIBentryALTinterwordspacing
G.~Blaj, ``Dead-time correction for spectroscopic photon counting pixel
  detectors,'' \emph{Journal of Synchrotron Radiation}, 2019, in press.
  [Online]. Available: \url{https://doi.org/10.1107/S1600577519007409}
\BIBentrySTDinterwordspacing

\bibitem{klavckova2019characterization}
\BIBentryALTinterwordspacing
I.~Kla{\v{c}}kov{\'a}, G.~Blaj, P.~Denes, A.~Dragone, S.~G{\"o}de, S.~Hauf,
  F.~Januschek, J.~Joseph, and M.~Kuster, ``Characterization of the {ePix100a}
  and the fastccd semiconductor detectors for the {European XFEL},''
  \emph{Journal of Instrumentation}, vol.~14, no.~01, p. C01008, 2019.
  [Online]. Available: \url{https://dx.doi.org/10.1088/1748-0221/14/01/C01008}
\BIBentrySTDinterwordspacing

\bibitem{blaj2016xray}
\BIBentryALTinterwordspacing
G.~Blaj, P.~Caragiulo, A.~Dragone, G.~Haller, J.~Hasi, C.~J. Kenney,
  M.~Kwiatkowski, B.~Markovic, J.~Segal, and A.~Tomada, ``X-ray imaging with
  {ePix100a}, a high-speed, high-resolution, low-noise camera,'' \emph{SPIE
  Proceedings}, vol. 9968, pp. 99\,680J--99\,680J--10, June 2016. [Online].
  Available: \url{https://dx.doi.org/10.1117/12.2238136}
\BIBentrySTDinterwordspacing

\bibitem{blaj2018performance}
\BIBentryALTinterwordspacing
G.~Blaj, A.~Dragone, C.~Kenney, F.~Abu-Nimeh, P.~Caragiulo, D.~Doering,
  M.~Kwiatkowski, B.~Markovic, J.~Pines, M.~Weaver, S.~Boutet, G.~Carini, C.-E.
  Chang, P.~Hart, J.~Hasi, M.~Hayes, R.~Herbst, J.~Koglin, K.~Nakahara,
  J.~Segal, and G.~Haller, ``Performance of {ePix10K}, a high dynamic range,
  gain auto-ranging pixel detector for {FELs},'' \emph{AIP Conference
  Proceedings}, vol. 2054, p. 060062, 2019. [Online]. Available:
  \url{https://doi.org/10.1063/1.5084693}
\BIBentrySTDinterwordspacing

\bibitem{blaj2018hammerhead}
\BIBentryALTinterwordspacing
G.~Blaj, D.~Bhogadi, C.-E. Chang, D.~Doering, C.~Kenney, T.~Kroll, J.~Segal,
  D.~Sokaras, and G.~Haller, ``Hammerhead, an ultrahigh resolution {ePix}
  camera for wavelength-dispersive spectrometers,'' \emph{AIP Conference
  Proceedings}, vol. 2054, p. 060037, 2019. [Online]. Available:
  \url{https://doi.org/10.1063/1.5084668}
\BIBentrySTDinterwordspacing

\end{thebibliography}
\end{document}